# Calculating Germinal Centre Reactions

Lisa Buchauer[1] and Hedda Wardemann[2]

[1] Division of Theoretical Systems Biology, German Cancer Research Center, 69120 Heidelberg, Germany
[2] Division of B Cell Immunology, German Cancer Research Center, 69120 Heidelberg, Germany

Corresponding authors: Lisa Buchauer, l.buchauer@dkfz-heidelberg.de and Hedda Wardemann, h.wardemann@dkfz-heidelberg.de

**Germinal centres are anatomically defined lymphoid organ structures that mediate B cell affinity maturation and affect the quality of humoral immune responses. Mathematical models based on differential equations or agent-based simulations have been widely used to deepen our understanding of the cellular and molecular processes characterizing these complex dynamic systems. Along with experimental studies, these tools have provided insights into the spatio-temporal behavior of B cells in germinal center reactions and the mechanisms underlying their clonal selection and cellular differentiation. More recently, mathematical models have been used to define key parameters that influence the quality of humoral vaccine responses such as vaccine composition and vaccination schedule. These studies generated testable predictions for the design of optimized immunization strategies.**

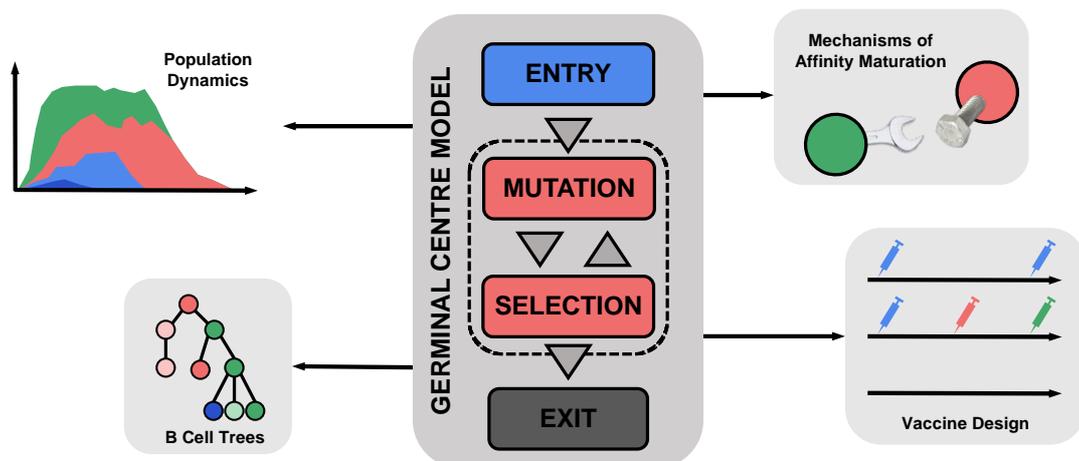

Graphical Abstract - Basic structure and applications of germinal centre simulations.

## Introduction

B cells are components of the adaptive immune system of higher vertebrates. Their main function is the production of soluble serum antibodies for neutralising pathogens. Each naïve B cell expresses a unique B cell antigen receptor (BCR), the membrane-bound form of its antibody, because the corresponding immunoglobulin (Ig) gene region is shaped by random genetic recombination during cellular development. This process alone yields a potential diversity of $10^{18}$ different BCR sequences in humans, largely exceeding the number of B cells in the body [1]. The diversity of the resulting B cell repertoire is mostly confined to the complementarity-determining regions (CDRs) of the antibodies, areas specialized for contacting and binding antigen, while the less diverse framework regions (FWRs) convey protein shape and stability.

Following infection or vaccination, those B cells whose BCRs bind to the foreign epitopes with above-threshold affinity are activated and recruited into the immune response. In the case of T cell-dependent antigens, activated B cells migrate to secondary lymphoid follicles where they seed germinal centres (GCs) [2–5]. In addition to proliferating B cells, these anatomically defined structures contain follicular dendritic cells (FDCs), stromal cells and T follicular helper (Tfh) cells. All three cell populations are essential mediators of affinity maturation, that is the improvement of humoral immunity with time and antigen exposure through the competitive selection of B cells with highest antigen affinity [6]. Serum antibodies influence ongoing GC reactions by antibody feedback, as stromal cells display antigen in the form of antibody-containing immune complexes and serum antibodies bind to FDC-displayed antigen in competition with GC B cells. Of note, GCs are unable to select for pathogen neutralisation, as antigenic epitopes are presented out of their functional context. GC selection is thus based on antibody-antigen affinity alone.

Within GCs, B cells actively mutate their immunoglobulin genes. This random somatic hypermutation process has the potential to generate B cell antigen receptors with improved affinity to their cognate antigen [7]. Ultimately, GC reactions improve humoral immune responses through the differentiation of GC B cells into short- and long-lived antibody-secreting plasma cells and the formation of memory B cells which can quickly respond to antigenic re-challenge [8,9].

Soon after the definitive experimental establishment of GCs as the sites of somatic hypermutation and affinity maturation based on rodent studies with model antigen [10,11], first mathematical models of the GC reaction were developed [12,13] as "data-interpretation tools" and to raise novel hypotheses [14]. Since then, experimental advances, particularly due to the use of novel *in vivo* imaging technologies [12,13], and model-based interpretation and formation of questions succeed each other in cycles; examples in which *in silico* exploration has preceded definite experimental validation include cycling of GC B cells between zones of mutation and selection [12,15], the existence of serum-antibody feedback in GCs [16,17] and the role of Tfh cells in GC selection [18,19].

In recent years, following breakthroughs in high-throughput single-cell Ig gene sequencing and expression cloning, large datasets have become available that allow monitoring B cell clonal diversification and expansion linked to BCR affinity [20–26]. This integrative approach opened a window to assess affinity maturation against real-world pathogens in humans during natural infection and in response to vaccination. In parallel, increasing numbers of vaccination-focused GC simulation studies have emerged. These integrate the available body of mostly mouse-derived knowledge and extrapolate to the human setting [27–30]. Such models aim at improving vaccine design in terms of vaccine composition and administration schedules.

Here, we review the most recent mathematical and computational models of the GC response, highlight interesting model features and discuss how they helped to improve our understanding of B cell responses specifically in vaccination settings. An overview of studies published since 2016 is given in Table 1 including information on the availability of the simulation codes. With this, we hope to facilitate the re-use and improvement of existing simulations as well as the development of new ones to address key questions in basic and applied immunology hand-in-hand with experimental data.

## Constructing theoretical GC models

Computational models of the GC reaction commonly take the form of agent-based stochastic simulations [31], systems of ordinary differential equations or combinations thereof. Alternatively, stochastic differential equations are employed. Independent of the framework used, core processes of the GC reaction must be cast into equations and computer-executable rules. As sketched in Figure 1, these include descriptions of cellular division, somatic hypermutation, competition and selection. Most models also formalize GC cell entry or seeding and GC exit via differentiation. Since affinity maturation via selection and mutation is intimately linked to the binding energy or affinity between antigen and antibodies, a set of rules yielding affinities between computational antigen and antibodies is required. Binding models are often designed with a specific research question in mind and have recently been reviewed in depth elsewhere [32].

Each of the core processes – activation, mutation, selection and differentiation – is influenced by a vast array of mechanisms, some of which are included in Figure 1. Choosing and assembling a meaningful subset of these puzzle pieces is an important step in the construction of a theoretical GC model. The choice of modules depends on the research question and positions the model on a spectrum with high resolution and small scale on one side and computational efficiency and larger scale on the other. On the extremes, one may aim to model a single GC comprised of diverse cell types at high spatial resolution [33,34] or a large ensemble of GCs jointly responding to several consecutive antigen exposures [24,35]. High-resolution models are important for figuring out the influence of each puzzle piece and thus for deciding which mechanisms to carry through to higher level models. They are, however, ill-suited to predict vaccination success at the organismal level as this builds on several sequential rounds of GC reactions, cross-talk between many GCs and the resulting amplification of rare high-affinity binders [24].

In general, computational GC models can only be as accurate as the underlying body of experimental knowledge. Open questions remain at all scales of GC modelling. For example, it is not known how the simultaneous presence of several antigens affects the response to individual antigens and whether the response resembles the sum of all parallel responses or caps at a defined maximum. Addressing these questions is complicated by differences in antigen and epitopes immunogenicity and stability. Furthermore, little is known about how the pre-vaccination B cell repertoire of individuals impacts the quality of their GC seeder cells and downstream vaccination outcomes. Model accuracy would further benefit from knowing how vaccine dosing influences the number of GC seeder cells, the diversity of GC Tfh cells and the size and number of resulting GCs.

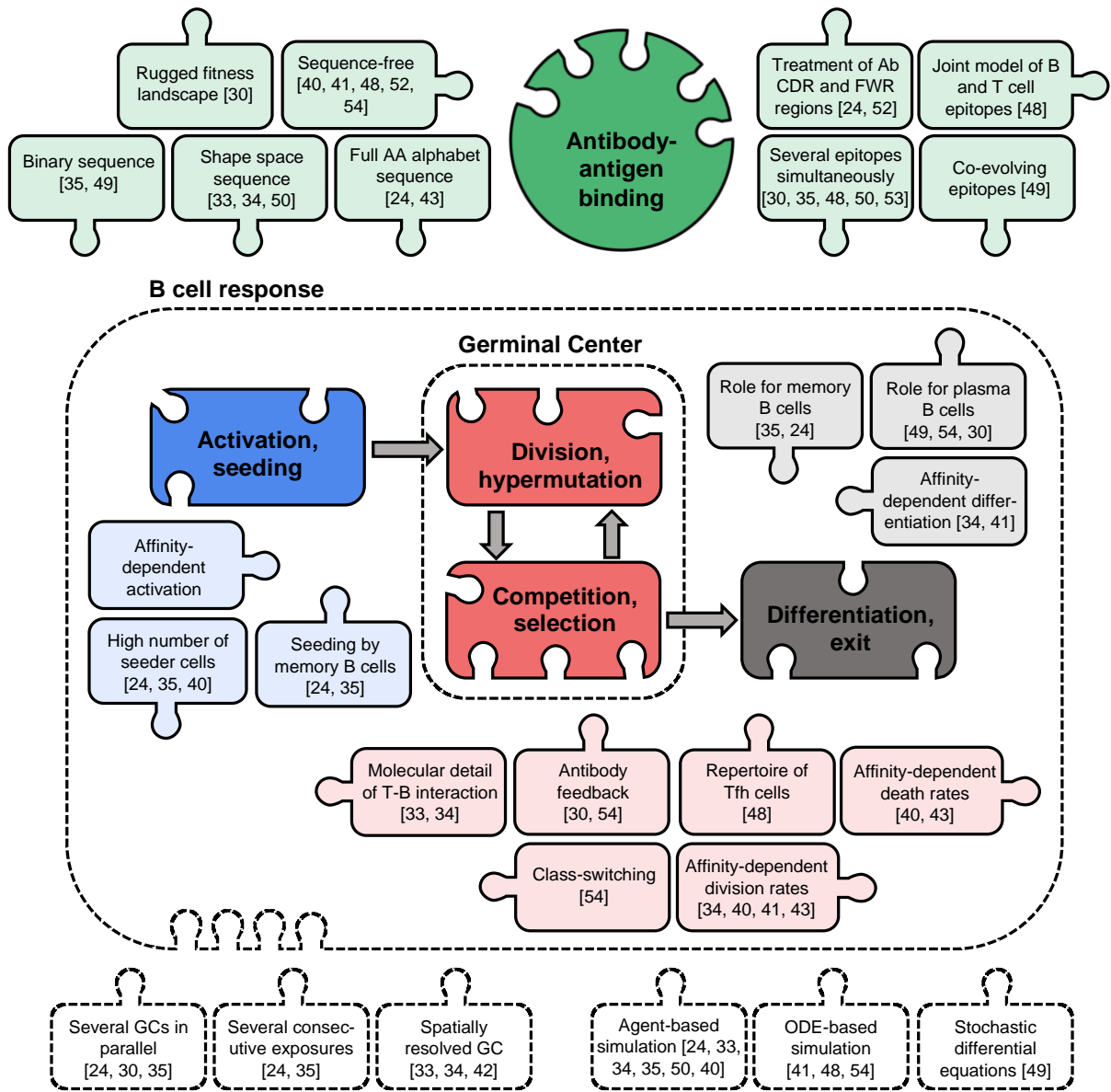

**Figure 1** – Modules of germinal centre models. Core modules (dark) are part of most simulations. Corresponding optional models (light) can be included depending on the research question and scale of interest. AA – amino acid, Ab – antibody, FWR – framework region, CDR – complementarity-determining region, ODE – ordinary differential equation, Tfh cell – T follicular helper cell, GC – germinal centre.

**Understanding clonal evolution in GCs**

GC B cell competition for limited resources and concomitant selective expansion rely on the fact that GCs are composed of B cells with different antigen affinities. The clonal composition changes throughout the GC lifetime as novel B cell clones enter, emerge by somatic mutation, expand and contract. Understanding these dynamics and the resulting clonal relationships of B cells found in the periphery is of central relevance when designing vaccines as it links to optimal epitope mixtures and scheduling.

Historically, it was believed that GCs are seeded by only 3-5 low-affinity founder cells which then expand massively while diversifying via somatic hypermutation [36,37]. However, recent *in vivo* visualisations of the murine GC response have revealed that individual GCs are typically seeded by 50-200 founder cells and lose diversity at different rates thereafter [38,39].

Nonetheless, many current simulations still build on the assumption that the number of founder cells is low. This is problematic as it restricts GC maturation in large to mutation-based affinity improvements. By contrast, the contribution of selective expansion of rare high-quality precursors to overall improvement can be substantial if a large and diverse pool of GC founder cells is considered [24]. Correspondingly, the mean affinity of a GC B cell population correlates negatively with its clonal diversity in a setting with many different founder cells [40]. However, highly abundant subclones do not necessarily have highest affinity, as best variants can arise late during the reaction and thus have insufficient time for expansion [41]. Caution is also necessary when interpreting the number of somatic hypermutations as an indicator of affinity maturation – the number of mutations an antibody has accumulated and its affinity do not necessarily correlate [24,41].

How clonal dynamics differ in settings with several, potentially differentially immunogenic epitopes present at the same time remains to be investigated experimentally. *In vivo* imaging tools like "brainbow GCs" [38] proved useful, but the interpretation of the data requires care – based on computer simulations several caveats regarding timing, dynamics and dosage of colour induction have been identified [42]. On a related note, GC models have been used to identify complications in the construction of B cell lineage trees with the aim of establishing mutational paths from germline Ig sequences to higher-affinity descendants. Here, simulations show that tree inference on selection-shaped datasets of BCR sequences is significantly more complicated than on datasets produced by neutral evolution [43].

**Manipulating GC selection processes**

Antibodies are correlates of protection in the majority of currently licensed vaccines and have been shown to also convey protection against HIV, malaria and influenza [44–46]. However, for several reasons including quickly mutating antigens and the parallel display of multiple epitopes, these diseases still evade efficient vaccination. GC simulations help to understand the challenges and aim to yield suggestions for vaccine compositions.

In this context, a current matter of investigation is how GCs, which are thought to be optimised for the selection of high-affinity B cells, are able to select for breadth, that is the ability to bind and ideally neutralize antigen variants. This is of particular importance for binding to quickly evolving pathogens such as HIV or influenza [46,47]. One possible mechanism conveying selection advantages can result from a diverse repertoire of Tfh cells: In contrast to their strain-specific competitors, broadly binding GC B cells take up several antigen variants with differing T cell epitopes and may therefore obtain help from a larger number of different Tfh cells [48]. It follows that vaccine mixtures intending to exploit this effect should ideally contain various different T and B cell epitopes. The resulting simultaneous presence of several epitopes may be comparable to the late stage of the natural HIV infection in which the viral population has strongly diversified. Here, the fixation probability of broadly binding lineages has been reported to exceed that of antibodies with higher specificity [49]. The optimal immune cocktail for eliciting broadly binding antibodies has been suggested to contain 6-8 antigen variants separated by 3-10 mutations [50]. However, others have argued that separating conflicting selection forces in time by sequential administration of variant antigens leads to more robust emergence of broadly binding antibodies [28,35].

Despite these computational results, it is unclear whether reliably eliciting broadly neutralizing antibodies against HIV by vaccination and thus within a practicable time frame is an attainable goal. Broadly neutralizing antibodies against HIV typically display high numbers of hypermutations many of which are located in the FWRs of antibodies [51]. The emergence of such sequences by the random process of hypermutations is an extremely rare event and consequently, on average, takes a long time. While existing computational models in this area rely on the assumption that a successful vaccination scheme exists, it would be equally relevant to computationally explore what the waiting time until emergence of broad binders is expected to be.

**Leveraging amplification of the right precursors**

It is often assumed that B cell immune responses start from low-affinity precursor cells only because pre-exposure serum titres are typically low. However, measuring serum titres is measuring a bulk quantity and a low titre does not exclude the existence of small populations of high-affinity B cells. As an example, in the case of the malaria parasite *Plasmodium falciparum*, previously unexposed human subjects were found to possess rare high-affinity naïve and memory B cells against the parasite surface protein PfCSP [24]. B cell vaccine responses relied largely on the clonal selection of these precursors along a triple-immunisation protocol. Based on a large-scale simulation comprising many individual GCs, an emerging avalanche effect in which a rare high-affinity precursor is first expanded in a single GC and its clonal descendants then go on to seed consecutively more GCs in secondary immune responses has been proposed as underlying mechanism. Activating the right precursors can increase immune response efficiency not only against malarial PfCSP, but also against HIV where it has been suggested to reduce the number of mutations required until the emergence of B cells expressing high-affinity broadly binding and neutralizing antibodies [52].

However, high-quality precursors are unlikely to exist against all antigens or in all patients. Here, novel vaccine administration schemes may help drive mutational affinity improvement by prolonging GC responses. While many established vaccines are administered 3-4 times at intervals of weeks to months with constant dosing [53], a recent mouse study tested various non-standard schemes including some with exponentially increasing or decreasing daily antigen doses over 1 or 2 weeks and employed mathematical modelling to interpret the results [54]. Serum-antibody feedback has been identified as a key factor. Early in the response, antigen is lost quickly from the lymph nodes, presumably because it cannot be retained in the form of immune complexes on FDCs by low-affinity antibodies. In contrast, antigen that is administered after a higher affinity response has developed can be captured and displayed for prolonged periods of time, thus allowing a much stronger response to develop. Such strategies have shown potential in non-human primate responses against HIV [55].

In summary, depending on boundary conditions like precursor availability and the mutational distance between them and a high-affinity binder, promising building blocks for vaccination strategies are emerging. Future simulations of the B cell response should pick up on this and aim at optimising protocols on a case-to-case basis. Considering the interplay of various germinal centres may prove to be essential for obtaining physiologically relevant results.

**Conclusions**

Computational models of affinity maturation can be employed to explore GC mechanisms as well as vaccination strategies. Interesting findings often emerge if such models are developed in conjunction with experimental studies, as mathematical models can assist the interpretation of experimental data and the development of novel testable hypotheses. Half of the studies reviewed here are directly available in online repositories (Table 1). This facilitates re-use and adaptation of existing models. In the future, computational models may mature to allow prediction of vaccination outcomes based on key parameters such as vaccine composition, administration schedule and the pre-vaccination B cell repertoire of individual patients.

Table 1: Recent Germinal Centre Simulations and Where to Find Them. Plain text versions of repository links are provided at the end of the manuscript.

| Topic | Motivation | Model Feature of interest | Language, Availability, Article |
|---|---|---|---|
| broadly binding antibodies (bbAbs) | Investigate how GCs can select for binding breadth | Uses an antigen model that consists of B and T cell epitopes, has explicit treatment of Tfh cell repertoire | C, available upon request [48] |
| | Facilitate eliciting bbAbs against HIV with immunization cocktails | Allows to vary composition of the vaccine: number of ingredients, mutational distance between them, individual concentration | *unknown*, no availability statement [50] |
| | Facilitate eliciting bbAbs in chronic infections like HIV | Analytical description of antigen-antibody co-evolution with accompanying simulation | Julia, available on github link [49] |
| | Understand evolutionary trajectories towards HIV-bbAbs from different starting points | Binding model with separate contributions of binding energies and antibody framework region flexibility | Python, available on github link [52] |
| GC dynamics & B cell diversity | Investigate how clonal diversity within individual GCs evolves over time | Implements and compares birth- and death-limited selection mechanisms | Matlab, available on github link [40] |
| | Generate benchmarking data for tree inference of antibody clones | Binding model uses nucleotide representation and replacements happen based on literature-backed mutabilities | Python, available on github link [43] |
| | Provide basis for interpretation of multi-colour stained GC reactions [38] | High-resolution 3D lattice simulation comprising Tfh cells, FDCs, stromal cells, B cells and their interactions | C++ and R, available on figshare link [42] |
| | Examine relation between clone abundance and affinity | Combines deterministic clonal evolution with stochastic emergence of new clones | R, available upon request [41] |
| molecular mechanisms | Understand the role of dopamine in Tfh-B cell contacts in GCs | Integrates specific molecular detail into existing agent-based model | C++, available upon request [33] |
| | Let GC processes emerge from cellular decisions rather than predefining event sequences | Links cellular behaviour and molecular mechanism via concentration-dependent rates | Matlab and Fortran 95, no availability statement [34] |
| vaccine schedules | Compare performance of sequential and parallel administration of HIV variants for eliciting breadth | Introduces binding model with conserved and variable parts as well as distracting epitopes | *unknown*, no availability statement [35] |
| | Understand dominance of lowly mutated, high-affinity B cells after malaria vaccination scheme | Disentangles contributions of selection and mutation to affinity maturation for simple and complex antigens | Python, available on github link [24] |
| | Understand antibody responses to several non-standard vaccination schemes | Incorporates antibody-feedback in a simple system of ODEs with predefined affinity maturation | (not applicable, small set of differential equations) [54] |

**Conflicts of Interest**

None.

**Acknowledgements**

Funding: LB was supported by the German-Israeli Helmholtz International Research School (Cancer-TRAX). The authors thank Verena Körber for proofreading the manuscript.

**Plain text addresses of code repositories in Table 1**

[49] https://github.com/jotwin/coevolution

[52] https://github.com/elifesciences-publications/paper-bnAb-flexibility/blob/master/am.py

[40] https://github.com/amitaiassaf/Modeling-Germinal-Center-Reaction

[43] http://doi.org/10.5281/zenodo.1306301

[42] https://figshare.com/articles/Computer_Simulation_of_MultiColor_Brainbow_Staining_and_Clonal_Evolution_of_B_Cells_in_Germinal_Centers/7068419

[24] http://doi.org/10.5281/zenodo.1048052